\documentclass[a4paper,11pt]{article}

\usepackage{jinstpub} 

\usepackage{graphicx}
\usepackage[table,xcdraw]{xcolor}
\usepackage{float}
\usepackage{mathtools}
\usepackage{hyperref}
\usepackage{amsmath}    
\usepackage{siunitx}
\usepackage{textgreek}
\usepackage[T1]{fontenc}
\usepackage[utf8]{inputenc}
\usepackage[italian]{babel}

\usepackage{lineno}

\title{\boldmath Cryogenic characterization of FBK NUV-HD-Cryo 3T SiPM sensors for the DUNE photon detection system}

\abstract{
The Deep Underground Neutrino Experiment (DUNE) is a long-baseline neutrino experiment based in the USA and composed of a Near Detector (ND) complex at Fermi National Laboratory (FNAL), and a Far Detector (FD) complex located at the Sanford Underground Research Facility (SURF) $\sim$1300\,km distant. DUNE will study neutrino oscillations looking for unresolved issues of the Standard Model of particle physics (SM) such as CP violation in the leptonic sector, neutrino mass ordering and others, starting from the early 2030s. The FD, with a mass of $\sim$17\,kt, that will exploit both ionization and scintillation signals to detect neutrino interactions with Argon. Scintillating photons in LAr will be detected by the photon detection system (PDS) based on light collectors coupled to Silicon Photomultipliers (SiPMs).
During a test campaign, different laboratories of the collaboration performed an investigation of the best SiPM candidates that fulfill the DUNE FD requirements. We identified two models of SiPM, produced by Hamamatsu Photonics K.K. (HPK) and Fondazione Bruno Kessler (FBK), respectively.
In this paper, we focus on the FBK selected model showing its main features. We will describe the characterization protocol, the results at both room and cryogenic temperatures and the photon detection efficiency measurements.
}


\author[1]{F. Acerbi}
\author[2]{M. Andreotti,}
\author[2,3]{A. Balboni,}
\author[4,5]{E. Bertolini,}
\author[6,7]{S. Bertolucci,}
\author[10,19]{G. Botogoske,}
\author[4,5]{F. Bramati,}
\author[4,5]{A. Branca,}
\author[4,5]{C. Brizzolari,}
\author[4,5]{G. Brunetti,}
\author[2,3]{R. Calabrese,}
\author[8]{E. Calvo,}
\author[10]{N. Canci,}
\author[4,5]{P. Carniti,}
\author[2,3]{D. Casazza,}
\author[4]{C. Cattadori,}
\author[9]{A. Cervera,}
\author[6,7]{F. Chiapponi,}
\author[2]{S. Chiozzi,}
\author[6]{V. Cicero,}
\author[2]{A. Cotta Ramusino,}
\author[4,5]{E. Cristaldo Morales,}
\author[8]{C. Cuesta,}
\author[2,3]{R. D'Amico,}
\author[6]{L. Degli Esposti,}
\author[4,5]{M. Delgado Gonzalez,}
\author[10,11]{F. Di Capua,}
\author[6]{D. Di Ferdinando,}
\author[12]{A. Dyshkant,}
\author[12]{M. Eads,}
\author[4,5]{A. Falcone,}
\author[13]{E. Fialova,}
\author[1]{A. Ficorella,}
\author[18]{P. Filip}
\author[10,11]{G. Fiorillo,}
\author[2,3]{M. Fiorini,}
\author[12]{K. Francis,}
\author[6,7]{A. Gabrielli,}
\author[4,5]{F. Galizzi,}
\author[14,15]{G. Gallina,}
\author[16]{D. Garcia-Gamez,}
\author[9]{M. \'A. Garc\'ia-Peris,}
\author[2,3]{T. Giammaria,}
\author[8]{I. Gil-Botella,}
\author[1]{A. Gola,}
\author[4]{C. Gotti,}
\author[2,3]{M. Guarise, \note{Corresponding author.}}
\author[4,5]{D. Guffanti,}
\author[6,7]{G. Ingratta,}
\author[6]{I. Lax,}
\author[8]{I. López de Rego,}
\author[2,3]{E. Luppi,}
\author[8]{S. Manthey,}
\author[9]{J. Martin-Albo,}
\author[6,7]{N. Mauri,}
\author[4,5]{L. Meazza,}
\author[6]{A. Mengarelli,}
\author[4,5]{A. Minotti,}
\author[6]{E. Montagna,}
\author[6]{A. Montanari,}
\author[2,3]{I. Neri,}
\author[16]{F. J. Nicolas-Arnaldos,}
\author[8]{C. Palomares,}
\author[1]{L. Parellada-Monreal,}
\author[6,7]{L. Pasqualini,}
\author[1]{G. Paternoster,}
\author[8]{L. P\'erez-Molina,}
\author[4]{G. Pessina,}
\author[6]{V. Pia,}
\author[2,3]{L. Pierini,}
\author[6]{F. Poppi,}
\author[6]{M. Pozzato,}
\author[9]{M. Querol,}
\author[15]{F. Retiere,}
\author[9]{J. Rocabado,}
\author[6,7]{A. Ruggeri,}
\author[9]{A. Saadana,}
\author[16]{A. Sanchez-Castillo,}
\author[16]{P. Sanchez-Lucas,}
\author[4,5]{A. Scanu,}
\author[2,3]{F. S. Schifano,}
\author[6]{G. Sirri,}
\author[13]{J. Smolik,}
\author[6]{M. Tenti,}
\author[4,5]{F. Terranova,}
\author[6]{V. Togo,}
\author[2,3]{L. Tomassetti,}
\author[4,5]{M. Torti,}
\author[6]{N. Tosi,}
\author[6]{C. Valieri,}
\author[8]{A. Verdugo de Osa,}
\author[17]{H. Vieira de Souza,}
\author[18]{J. Zalesak,}
\author[16]{B. Zamorano,}
\author[6,7]{S. Zucchelli,}
\author[12]{V. Zutshi,}

\affiliation[1]{Fondazione Bruno Kessler, 38123 Trento, Italy}
\affiliation[2]{Istituto Nazionale di Fisica Nucleare, Sezione di Ferrara, I-44122 Ferrara, Italy}
\affiliation[3]{Universit\`a degli Studi di Ferrara, 44122 Ferrara, Italy}
\affiliation[4]{Istituto Nazionale di Fisica Nucleare Sezione Milano Bicocca}
\affiliation[5]{Università di Milano Bicocca, Dipartimento di Fisica}
\affiliation[6]{Istituto Nazionale di Fisica Nucleare Sezione di Bologna, 40127 Bologna, Italy}
\affiliation[7]{Università di Bologna, 40127 Bologna, Italy}
\affiliation[8]{CIEMAT, Centro de Investigaciones Energéticas, Medioambientales y Tecnológicas, E-28040 Madrid,}
\affiliation[9]{Instituto de Física Corpuscular, Valencia, Spain}
\affiliation[10]{Istituto Nazionale di Fisica Nucleare Sezione di Napoli, Naples, Italy}
\affiliation[11]{Università degli Studi di Napoli Federico II}
\affiliation[12]{Northern Illinois University, Department of Physics}		
\affiliation[13]{Czech Technical University, 115 19 Prague 1, Czech Republic}
\affiliation[14]{Physics Department, Princeton University, Princeton, NJ 08544, USA}
\affiliation[15]{TRIUMF, 4004 Wesbrook Mall, Vancouver, BC V6T 2A3, Canada}
\affiliation[16]{University of Granada \& CAFPE, Campus Fuentenueva (Edif. Mecenas), 18002 Granada, Spain}
\affiliation[17]{APC, Laboratoire Astroparticule et Cosmologie, Université de Paris Cité, Paris, France}
\affiliation[18]{Institute of Physics, Czech Academy of Sciences, 182 00 Prague 8, Czech Republic}
\affiliation[19]{Università degli Studi di Padova, Dipartimento di Fisica e Astronomia}

\emailAdd{marco.guarise@fe.infn.it}

\begin{document}
\maketitle
\flushbottom

\section{Introduction}
\label{sec:intro}


The Deep Underground Neutrino Experiment (DUNE) is a next-generation long-baseline neutrino experiment whose main goal will be a detailed study of neutrino oscillation. DUNE will consist of a Near Detector (ND) placed in proximity of the neutrino production site at Fermi National Laboratory  to characterize the neutrino flux and interaction, and of a Far Detector (FD) that will be installed at Surf Underground Laboratory in South Dakota~\cite{DUNE:2020lwj} to measure the oscillated neutrino spectrum. Of the first two Far Detector modules, one will be a liquid argon time projection chamber (LAr TPC) with an Horizontal Drift (HD) configuration and a total mass of nearly 17~kt. Beyond charge signals, DUNE FD-HD will also exploit the scintillation light of argon whose peak wavelength is at $\sim$127 nm~\cite{Heindl_2010}. Light signals will be collected in the Photon Detection System (PDS) thanks to the so-called X-ARAPUCA modules. These modules allow to trap photons inside a highly reflective box that contains wavelength shifting bars and visible-light sensitive Silicon Photo-Multipliers (SiPM) sensors~\cite{Souza_2021}.

FD-HD will employ SiPMs produced by two different companies: Fondazione Bruno Kessler (FBK) and Hamamatsu Photonics K.K (HPK). Both models are designed for their use at cryogenic temperatures and have an effective area of around 36 mm$^2$, while they have different characteristics in terms of cell pitch, breakdown voltage and quenching resistor. These models were selected after an R\&D phase, carried out by the DUNE PDS Consortium in collaboration with the manufacturers, looking for the prototypes that addressed the technical requirements for the FD-HD\,\cite{DUNE:2021hwx, andreotti2024cryogenic} and are thus "within DUNE specifications". These requirements are listed in table\,\ref{tab:specs}. Regarding FBK sensors, a total of 250 sensors were produced for this purpose. The tests concerning FBK NUV-HD-cryo 3T sensors, that will be presented in this paper, have been performed in seven different test sites of the PDS Consortium following the same methodology and procedure used for the tests previously conducted on HPK SiPMs~\cite{andreotti2024cryogenic}. The laboratories are the following: Bologna (\textit{Istituto Nazionale di Fisica Nucleare} and \textit{Università di Bologna}), DeKalb (\textit{Northern Illinois University, Department of Physics}), Ferrara (\textit{Istituto Nazionale di Fisica Nucleare} and \textit{Università di Ferrara}), Madrid (\textit{CIEMAT, Centro de Investigaciones Energ\'eticas, Medioambientales y Tecnol\'ogicas}), Milano (\textit{Istituto Nazionale di Fisica Nucleare} and \textit{Università di Milano-Bicocca}), Napoli (\textit{Istituto Nazionale di Fisica Nucleare} and \textit{Università di Napoli}), Prague (\textit{Institute of Physics, Czech Academy of Sciences}), and Valencia (\textit{Instituto de Física Corpuscular}).

\begin{table}
\centering
 \begin{tabular}{|c|c|c|} 
   \hline
   \textbf{Parameter} & \textbf{Value} & \textbf{Note} \\
   \hline
   SiPM dimensions  & 6x6 mm$^{2}$ & Compatible with the PDS  \\
                       & &  modules   \\
   \hline
   Cell pitch & 50-150\,$\mu$m &  \\
   \hline
   PDE  &  $>$35\% at 430\,nm & At room temperature  \\
    &                        & and @ nominal voltage (Vop)\\
   \hline
   Window material & Siliconic or epoxidic & Cryogenic reliable\\
   \hline
   DCR & $<$200 mHz/mm² at Vop & @ 77\,K \\
   \hline
   Cross-talk probability & $<$35\% at Vop & @ 77\,K\\
   \hline
   After-pulsing probability & $<$5\% at Vop &  @ 77\,K \\
   \hline
   SiPM recovery time & 200-1000\,ns & Optimal for cold electronics \\
   \hline
   Breakdown voltage (Vbd) spread & $<$200\,mV (max-min) &In a PDS module \\
   
   \hline
   Maximum Vbd voltage spread & $<$2\,V (max-min) & Global spread\\
   \hline
   Resilience to  & $>$20 & Tested at 77\,K \\
              thermal cycles    &        & by the Consortium \\                                  
   
   \hline
\end{tabular}
\caption{Specific Requirements for the SiPMs of the  FD1-HD Photon Detection System.}
\label{tab:specs}
\end{table}


The paper is structured as follows: Section\,\ref{sec:SiPM} presents a comprehensive overview of the key features of FBK sensors specifically customized for DUNE, in comparison to other FBK models. Section\,\ref{sec:characterization}  details the measurements performed by members of the PDS consortium to characterize the SiPMs at cryogenic temperatures, including the experimental setup and procedures used. Section\,\ref{sec:PDE_meas}  addresses the photon detection efficiency (PDE) measurements. Finally, Section\,\ref{sec:conclusion} discusses the results and concludes the study.

\section{SiPM features}
\label{sec:SiPM}


\subsection{NUV-HD-Cryo Technology}
\label{sec:NUV-HD-Cryo}

Over the years, FBK has developed various technologies of SiPMs to obtain optimal performances in order to meet the requirements of different experiments and applications. SiPMs are matrices of Single-photon avalanche diodes (SPADs) each one with an integrated quenching resistor, separated by a non-active trench and connected together in parallel forming a single sensor. Depending on the type of epitaxial layer used for the fabrication, we can distinguish two families of technologies: the first, is a p-on-n junction with an n-type epitaxial layer labeled as FBK NUV-HD/VUV-HD. This technology is characterized by a peak detection efficiency in the near-UV\,\cite{gola2019nuv}. The second technology is n-on-p junction with a p-type epitaxial layer named FBK RGB/NIR-HD, featuring a peak detection efficiency at longer wavelengths.
Starting from the standard NUV-HD technology, a customization of the electric field was implemented to develop the NUV-HD-Cryo SiPM technology for cryogenic applications such as in the DarkSide-20k experiment\,\cite{darkside24}. Strengths of this technology are a very low dark count rate (DCR) in the order of few mHz/mm$^2$ at cryogenic temperature and a lower afterpulsing probability (AP), as a result of a low peak value of the electric field compared to the standard FBK technology\,\cite{acerbi2017cryogenic}.

\subsection{DUNE Customization}
\label{sec:DUNE_SiPM}

In the framework of the DUNE experiment, a dedicated customization of the NUV-HD-Cryo SiPM technology was carried out to improve the performance of the detectors to better fit the requirements of the experiment.
The gain of a SiPM microcell, defined as the number of carriers generated during an avalanche process, is a key factor to  improve the signal to noise ratio enhancing the output signal. During the avalanche build-up the number of generated carriers is proportional to the active volume of the microcell, thus increasing the cell size the gain increases. However, this gain increment comes with a drawback of a higher optical crosstalk (CT), because the generation of secondary photons is proportional to the amount of carrier flowing during the avalanche. The aim of the DUNE customization was to modify the standard SiPM structure to obtain a device that combines a high gain and a limited CT. This was done by increasing the Deep Trench Isolations (DTIs) between neighboring cells. 
Two different layout splits were produced and tested to evaluate the best option for the DUNE experiment, a standard single DTI (called 1-trench or 1T) with 30 \textmu m cell pitch device to be used as a reference, and a 54 \textmu m cell pitch device with three DTIs (hereinafter referred to as triple trench or 3T).

\begin{figure}
\centering
  \includegraphics[width=\textwidth]{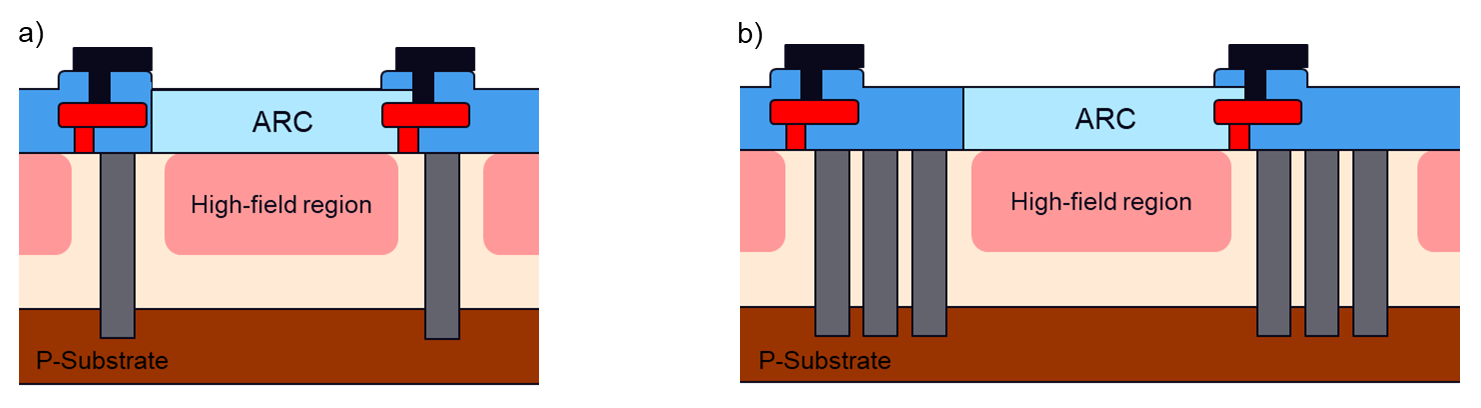}
  \caption{a)	Cross-section of the 30 \textmu m cell pitch device with a standard single DTI (1T – 30 \textmu m), b) Cross-section of the 54 \textmu m cell pitch device with three DTIs (3T – 54 \textmu m). High-field region in pink, quenching resistor in red, metal contact in black, passivation layer in blue, Active Reflective Coating in light blue and Deep Trench Isolation in gray.}
  \label{fig:FBK_cross-section}
\end{figure}

In figure\,\ref{fig:FBK_cross-section} the cross-sections of the two different layout splits with all the basic features of the cell structure are depicted. The high-field region (pink structures in the figure) where the avalanche takes place is implemented by a high energy ion implantation. The quenching resistors, which are needed to quench and subsequently recharge the microcells, are represented in red, while the metal contacts are shown in black. On the top part of the cell, a passivation layer is represented in blue, while the Active Reflective Coating (ARC), designed to minimize the light reflection in the visible spectral range, is depicted in light blue. Each cell of the SiPM is separated by deep trenches (gray structures) filled with silicon dioxide that are effective to isolate the cells electrically, and also partially optically from photons from nearby cells.

\subsection{Features at room temperature}
\label{sec:FBK_Characterization}

\begin{figure}
\centering
  \includegraphics[width=0.6\textwidth]{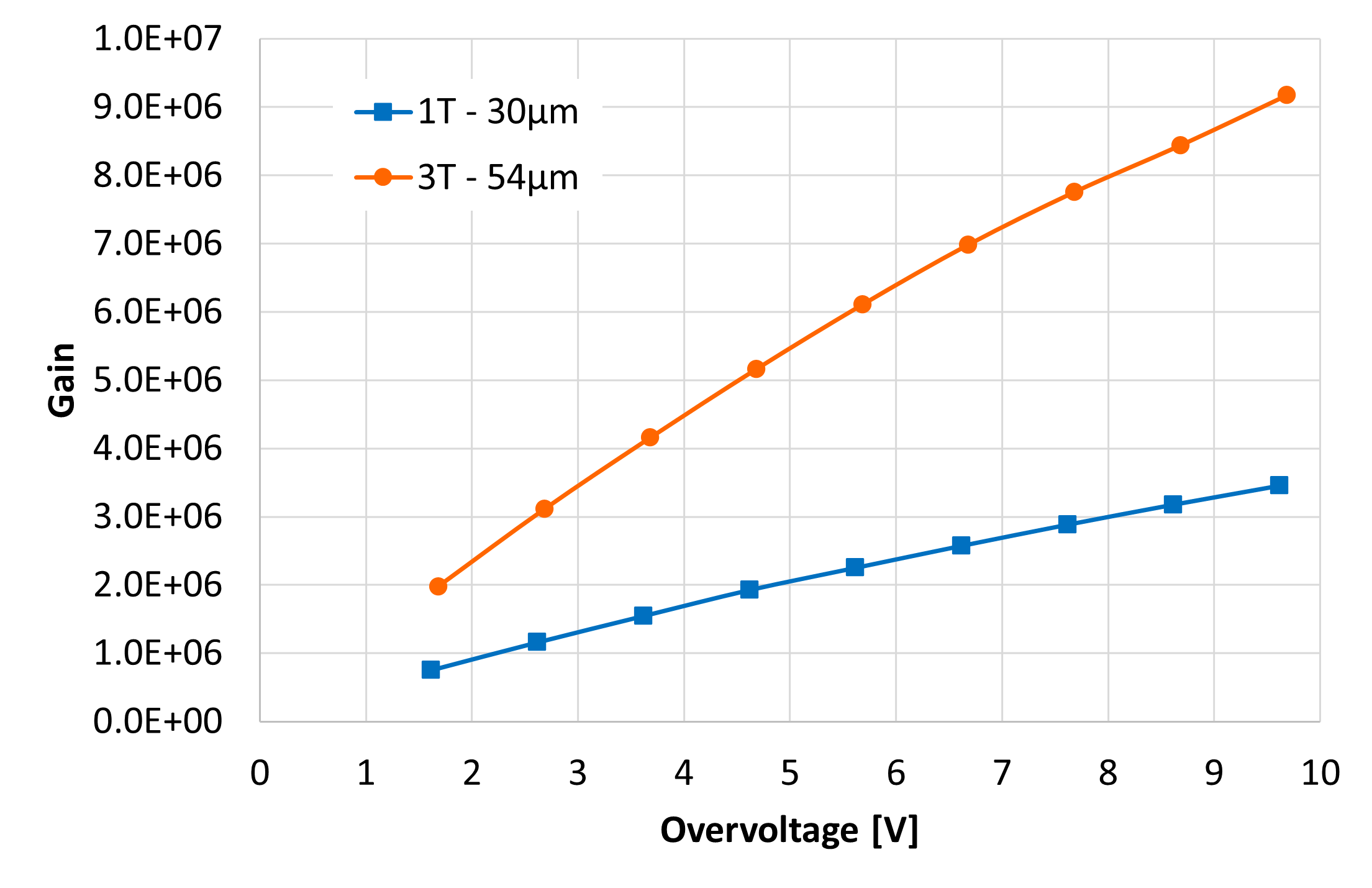}
  \caption{Comparison of the Gain as a function of the overvoltage between the two different splits (1T – 30 \textmu m; 3T – 54 \textmu m). Error bars lower than 1\% not visible in the graph.}
  \label{fig:FBK_Gain}
\end{figure}

This section presents the results of a functional characterization of the two layout splits, providing a comparison in terms of gain, CT and PDE. The characterization was carried out at the FBK laboratory immediately after the production of the splits.
Measurements were acquired by connecting the devices to a trans-impedance amplifier with a total gain of 5000 V/A, considering the 50 \si{\ohm} termination of the oscilloscope input, at a stable 20 degrees Celsius temperature in a climatic chamber. The gain of the devices is calculated from the single-cell area of the acquired waveform knowing the gain of the amplifier. A detailed procedure for the characterization of SiPMs is described in a dedicated paper\,\cite{piemonte2012development}. The PDE was measured at room temperature by using a pulsed mode technique with a setup featuring an integration sphere and LEDs of different wavelength, for further details refers to\,\cite{zappala11set}.
In figure\,\ref{fig:FBK_Gain} the gain as a function of the overvoltage is shown. The gain of the triple trench is approximately 2.6 times the gain of the single trench split. This value is very close to the expected value (2.8), calculated considering the geometrical ratio between the total area of the microcells excluding the trenches region. The difference between the expected and measured value of the gain can be attributed to border effects, related to the virtual guard ring around the active area of the microcells, which is characterized by higher depletion and, thus, lower capacitance per unit area.

\begin{figure}
\centering
  \includegraphics[width=0.6\textwidth]{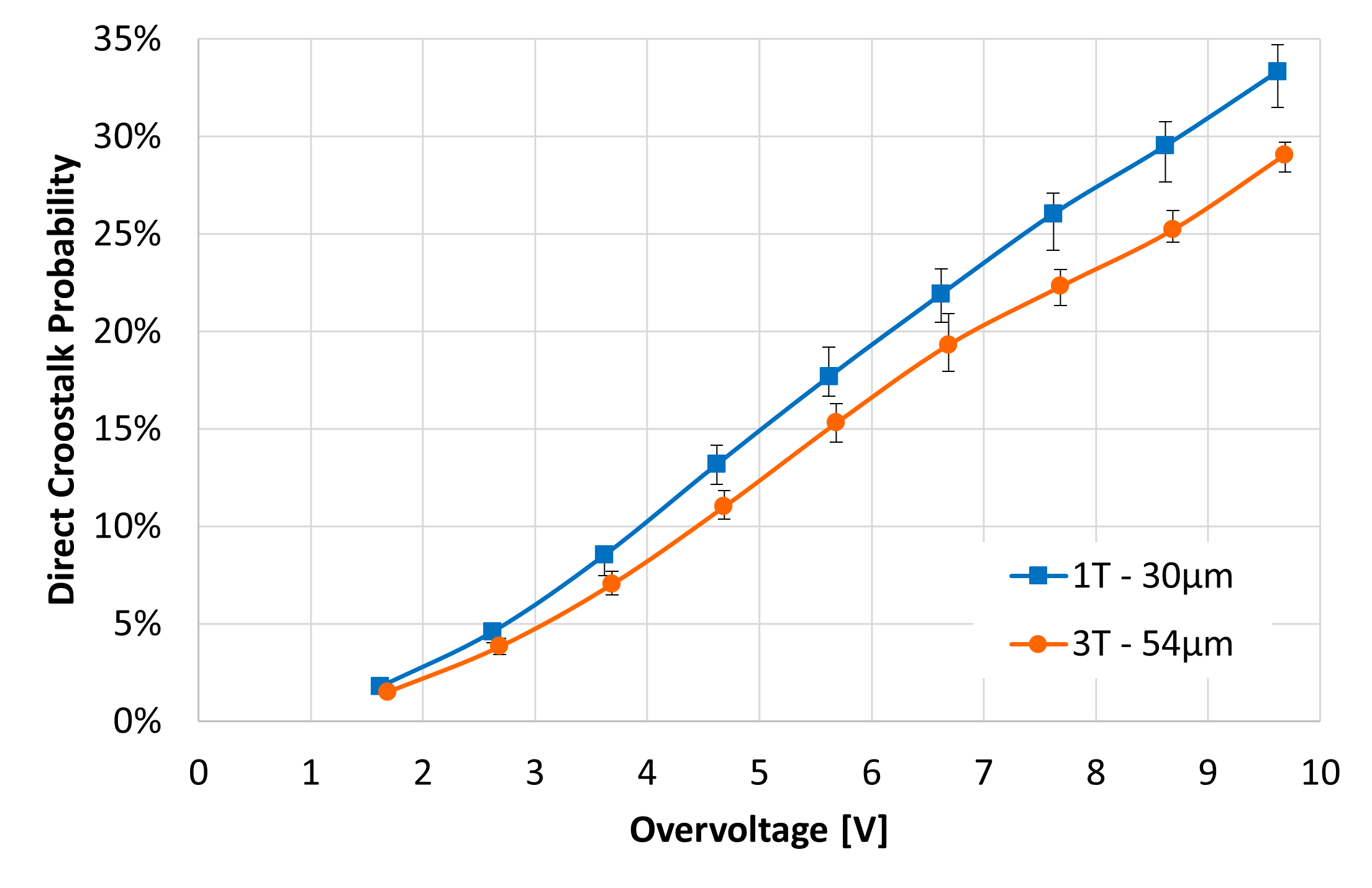}
  \caption{Comparison of the direct CT probability as a function of the overvoltage between the two different splits (1T – 30 \textmu m; 3T – 54 \textmu m).}
  \label{fig:FBK_CT}
\end{figure}

In figure\,\ref{fig:FBK_CT} the CT probability as a function of the overvoltage is reported. The single trench layout exhibits a slightly higher value with respect to the triple trench. The reduction of the CT for the triple trench devices is significant considering the 2.6 times higher gain. Measurements were performed on test structures of 1x1\,mm$^2$ active area without any protection resin on top of the devices.

\begin{figure}
\centering
  \includegraphics[width=0.6\textwidth]{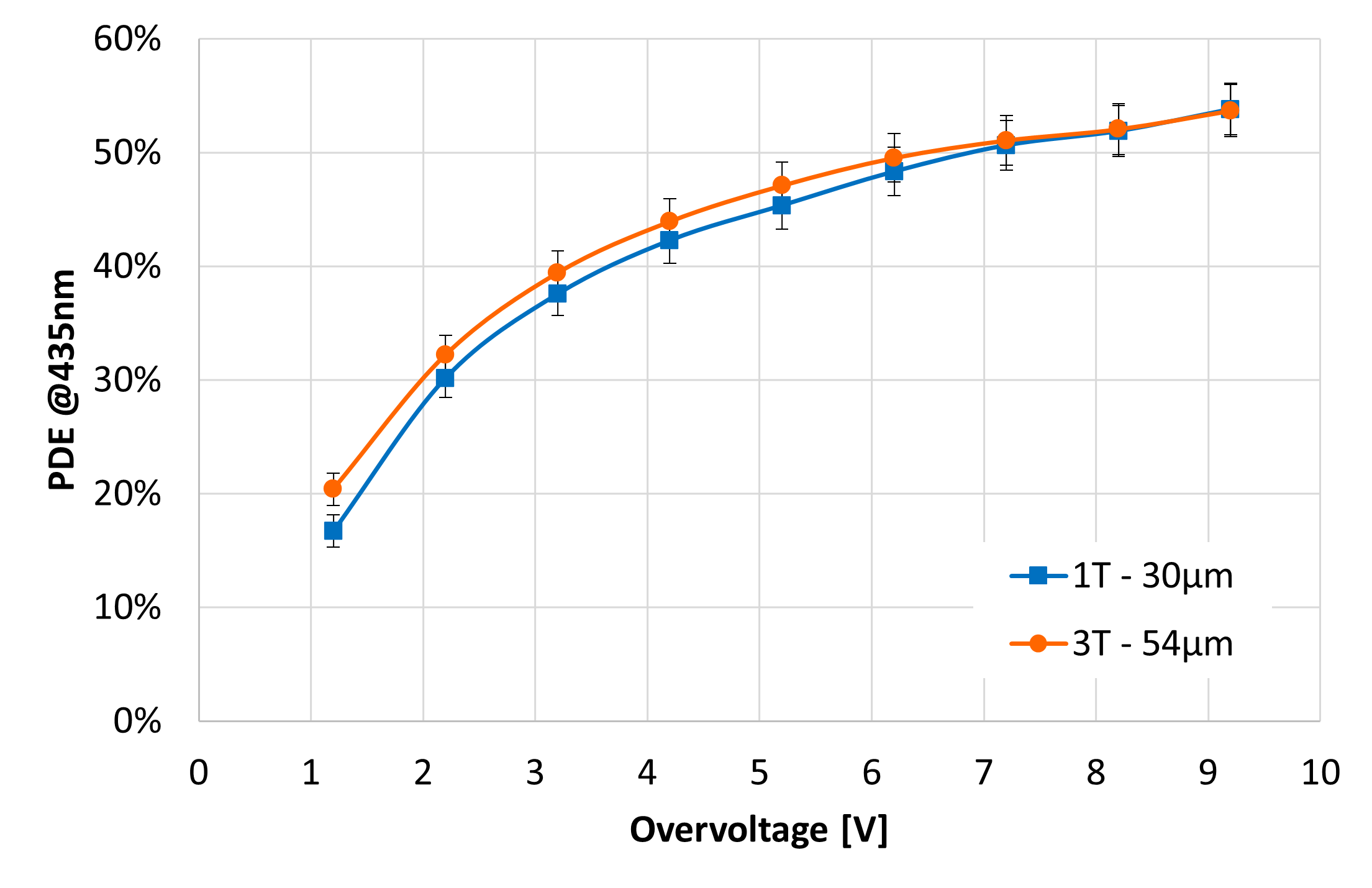}
  \caption{Comparison of the PDE at room temperature and 435\,nm as a function of the overvoltage between the two different splits (1T – 30 \textmu m; 3T – 54 \textmu m).}
  \label{fig:FBK_PDE}
\end{figure}

The PDE at 435 nm as a function of the overvoltage is shown in figure \ref{fig:FBK_PDE}. Measurements were performed on a test structure of 1x1\,mm$^2$ active area with an epoxy resin layer of approximately 500 \textmu m on top of the devices to smooth the oscillations in the PDE spectrum created by the constructive and destructive interferences caused by the top anti reflection coating on the SiPM surface. 
PDE increases with increasing overvoltage thanks to the increased triggering probability, i.e. the probability for a primary carrier to trigger an avalanche, reaching a saturation level at higher overvoltage \cite{zappala2016study}. The two different splits exhibit similar values as expected from the similar geometrical fill factor between the two layouts (75\% and 77\%).

As a conclusion of this section, the two different layout splits exhibit similar CT and PDE while the gain of the triple trench is around 2.6 times the gain of the single trench split independently of the overvoltage. The 54 \textmu m-cell pitch device with three DTIs was selected for the DUNE experiment for its higher gain.

\section{Cryogenic characterization}
\label{sec:characterization}

\subsection{Experimental apparatus and procedure}
\label{sec:setup}



As previously described in section\,\ref{sec:intro}, different institutions belonging to the DUNE PDS consortium were involved in the characterization of the sensors. To ensure consistent and comparable results across all laboratories, each testing apparatus adheres to unified specifications established by the DUNE PDS consortium. This section provides a thorough description of the prototype setup and the detailed protocol followed during measurements. Each laboratory implemented similar configurations meeting the same or superior specifications, thereby maintaining methodological consistency and reproducibility.


\begin{figure}
\centering
  \includegraphics[width=\textwidth]{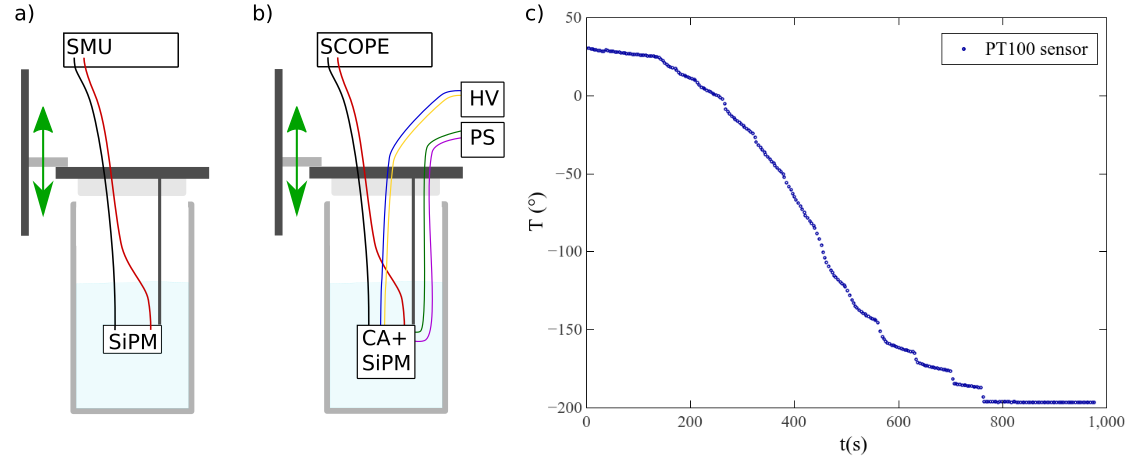}
  \caption{Setup used for the characterization of the sensors. a) Sketch of the apparatus used for IV measurements. b) Sketch of the apparatus used for DCR study, c) Temperature of the PT100 sensor placed near the SiPM as a function of time during the diving in LN2 phase.}
  \label{fig:setup}
\end{figure}

The setup consists of a liquid nitrogen (LN2) dewar where the sensors are placed in order to perform tests at cryogenic temperatures. Depending on the type of measurements to perform, different commercial instruments, with different features, were used. Sketches of the apparatus are shown in figures\,\ref{fig:setup}-a and \ref{fig:setup}-b.
The apparatus is also instrumented with a mechanical linear stage that allows to dive the sensors into LN2 following a controlled thermal profile. This can be done thanks to a temperature sensor (PT100) placed in the proximity of the SiPM. An example of a typical thermal profile during the diving phase is shown in figure\,\ref{fig:setup}-c.

The SiPMs current-voltage characteristic curve (IV) measurements were performed by a commercial source meter unit (SMU) directly connected to the SiPM through triaxial cables while the DCR characterization required a more complex apparatus that exploit the SMU to bias the SiPM with a stabilized voltage, a cryogenic amplification (CA) stage and an oscilloscope to acquire the signals. 
For the IV measurements, the minimum requirements for the SMU are (0-60)\,V voltage range, 10\,pA precision, 6 digits resolution used with cables whose loss is less than 10\,pA. 

For the DCR and gain measurements, the amplification stage is common for all the labs and is described in a dedicated publication\,\cite{brizzolari2022cryogenic}. The output of the amplifier is connected to the input of a digital oscilloscope, whose minimum requirements, established by the PDS consortium, are 1\,GHz bandwidth, 5\,Gs/s and 8\,bit resolution. The DCR measurement is performed in a completely dark environment at three different values of overvoltage ($V_\mathrm{OV}$), acquiring the waveforms with a trigger threshold of 0.5 photoelectron. We perform a dedicated analysis based on single signals observing their temporal and amplitude distributions. Integrating the SiPMs pulse signals, we also estimated their gain. 

The measurements sequence has also been defined to be common to all the labs involved in the test characterization. It was organized as follows:
\begin{itemize}
    \item IV curve at room temperature;
    \item 1$^{st}$ diving phase (half thermal cycle);
    \item IV curve at LN2 (before thermal cycles);
    \item emerging phase (half thermal cycle);
    \item 18 thermal cycles (room temperature--LN2--room temperature);
    \item diving phase (half thermal cycle);
    \item IV curve at LN2 (after thermal cycles);
    \item emerging phase (half thermal cycle);
    \item diving phase (half thermal cycle);
    \item DCR and gain at three different values of $V_{OV}$;
    \item emerging phase (half thermal cycle).
\end{itemize}

Each sensor is thus subjected at an overall number of 20 thermal cycles (diving phase from room temperature to LN2 and emerging phase from LN2 to room temperature).
Finally, the results from the different laboratories are compared and organized together to obtain common results that will be presented in this publication in section\,\ref{sec:results}.

\subsection{Cryogenic characterization results}
\label{sec:results}

\begin{figure}
\centering
  \includegraphics[width=\textwidth]{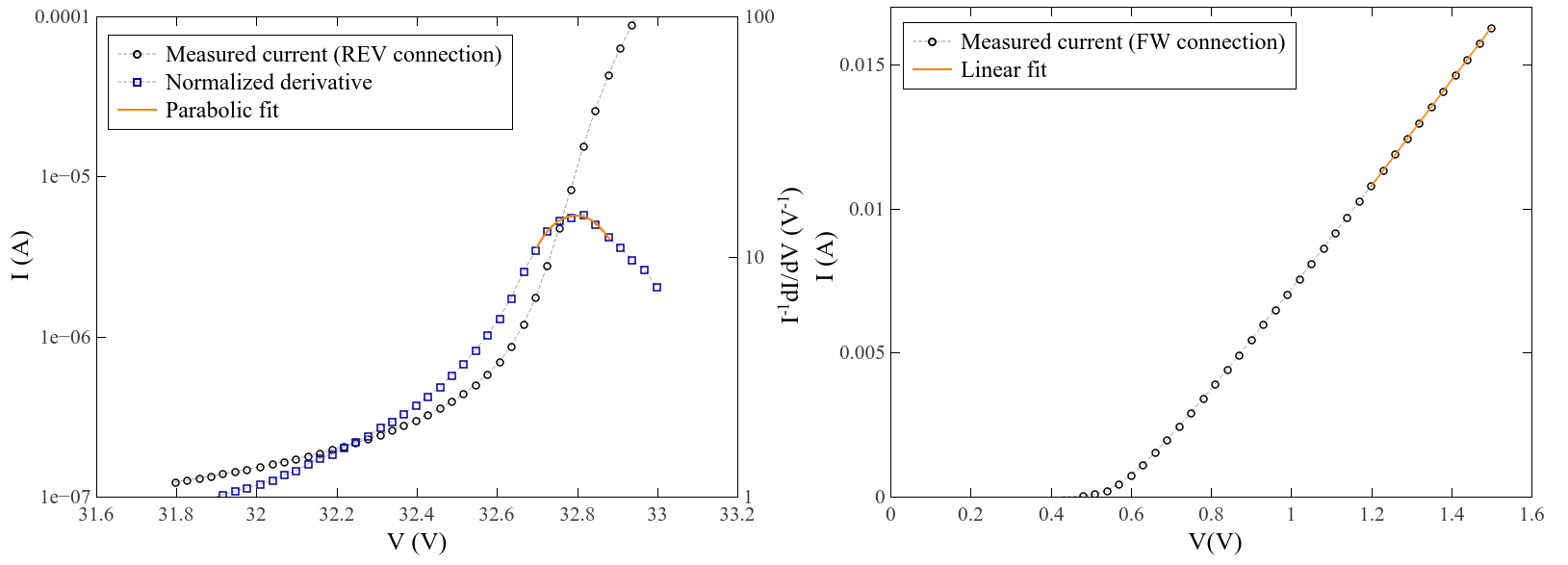}
  \caption{Left side: IV curve in reverse mode (REV) measured at room temperature (black circles). The normalized derivative (blue squares) and the parabolic fit are also shown in the plot. Right side: IV curve in the forward (FW) region measured at room temperature (black circles). A linear fit is also shown in the plot.}
  \label{fig:IVcurve}
\end{figure}

With the setup described in section \ref{sec:setup}, the different laboratories performed the measurements of the main features of the FBK NUV-HD-Cryo 3T SiPMs. 
The IV reverse and forward curves allowed us to obtain the breakdown voltage ($V_\mathrm{BD}$) and the quenching resistor ($R_\mathrm{q}$) respectively. The value of $V_\mathrm{BD}$ was estimated looking at the peak of the normalized derivative of the IV curve in the reverse region
. The maximum of the curve $I^{-1} dI/dV $ was fitted with a 2\textsuperscript{nd} degree polynomial function in order to estimate the best value of $V_\mathrm{BD}$.
In the forward region of the IV curve, a linear fit in the range [1.2-1.5]\,V allowed us to get the value of the global quenching resistor of the sensor ($R_\mathrm{SiPM}$).
The value of the quenching resistor of each cell ($R_\mathrm{q}$) can be then easily derived multiplying by the number of cells (assuming equal resistors for each cell).
Figure\,\ref{fig:IVcurve} shows two examples of the measured IV and the fits in the forward and reverse regions while SiPM is kept at room temperature.

Table\,\ref{tab:IV} summarizes the main parameters obtained during IV measurements at room and at LN2 temperatures after the 19$^{th}$ thermal cycle.

\begin{table}[h!]
    \centering
    \begin{tabular}{|c|c|c|}
    \hline
    \textbf{Temperature} &  \textbf{$V_{bd}$ (V)} & \textbf{$R_q (k\Omega)$} \\
    \hline 
    300\,K   &33.0$\pm0.1$  & 560$\pm10$ \\
    \hline
    77\,K & 27.08$\pm0.03$ & 2640$\pm70$\\
    \hline
    \end{tabular}
    \caption{Mean values and standard deviations for $V_\mathrm{BD}$ and $R_q$ obtained from the various laboratories both at room temperature and at LN2 temperature after the 19$^{th}$ thermal cycle.}
    \label{tab:IV}
\end{table}

\begin{figure}
\centering
  \includegraphics[width=\textwidth]{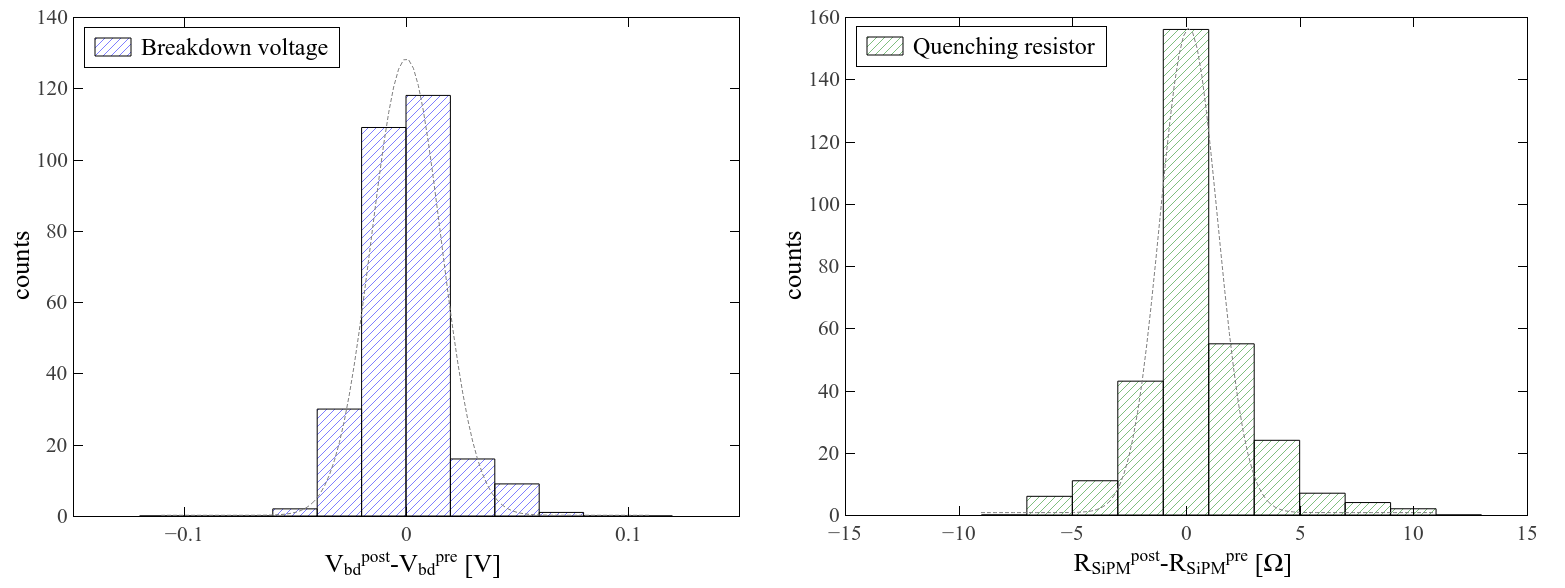}
  \caption{On the left part: histogram of the differences in the breakdown voltage values measured before and after the thermal cycles fitted with a Gaussian distribution with a mean value of 0\,V and a standard deviation of 0.03\,V. On the right figure: histogram of the differences in the quenching resistor values measured before and after the thermal cycles with superimposed a Gaussian distribution with mean value of 0.1\,$\Omega$ and a standard deviation of 2.5\,$\Omega$.}
  \label{fig:comparison_prepost}
\end{figure}

In order to check the resilience of the sensors to the thermal stresses, we also checked on every sensor the variation of both breakdown voltage and quenching resistor before and after the thermal cycles. Figure\,\ref{fig:comparison_prepost} shows on the left the histogram for the breakdown voltage difference where $V_\mathrm{BD}^\mathrm{pre}$ and $V_\mathrm{BD}^\mathrm{post}$ represent respectively the value at the 1$^{st}$ and at the 19$^{th}$ thermal cycle. The distribution is centered at 0\,V as expected if no damages occur, and has a width calculated as the standard deviation of the Gaussian distribution of (31$\pm$2)\,mV. On the right part, the histogram of the difference between the entire SiPM quenching resistance (as calculated from the forward IV curve) after the thermal stresses ($R_\mathrm{SiPM}^\mathrm{post}$) and before the thermal clycling ($R_\mathrm{SiPM}^\mathrm{pre}$) is shown instead. This last distribution is centered at 0\,$\Omega$ and it has a standard deviation of ($2.5\pm0.1)\,\Omega$ which is negligible, considering that the systematic uncertainty in the measurements of the resistor due to the electrical contacts at room and LN2 temperature can be estimated as few Ohms.

 No values differing by more than 70\,mV in $V_\mathrm{BD}$ and 10\,$\Omega$ in $R_q$ (in absolute value) were found, indicating a good reproducibility of the measurements and no substantial variations in the SiPM main characteristics after thermal stresses.

\begin{figure}
\centering
  \includegraphics[width=\textwidth]{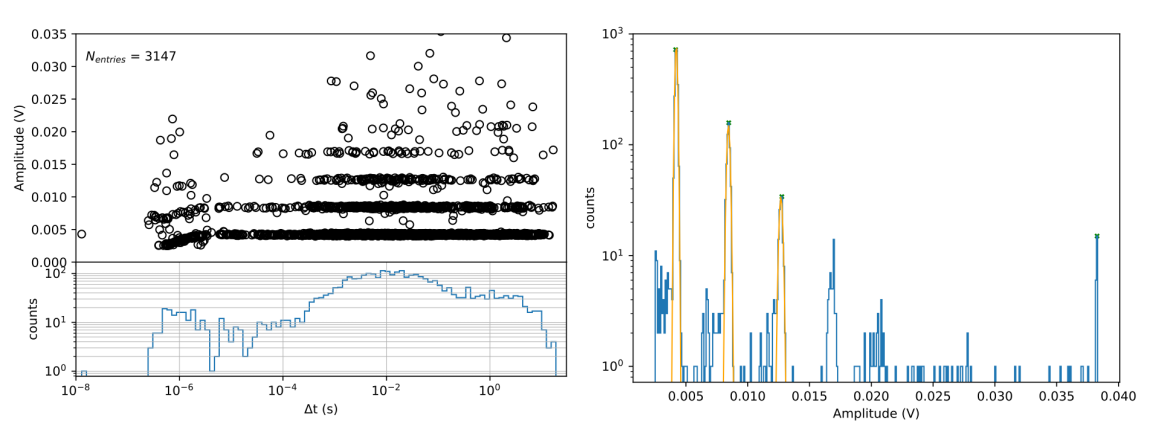}
  \caption{On the left side, an example of a scatter plot of amplitude versus time delay between consecutive events (top) and the corresponding x-axis projection histogram (bottom). On the right side the amplitude histogram of single signals in logaritmic-scale is shown, it corresponds to the y-axis projection of the figure on the left. Yellow curves superimposed to the distribution represent the Gaussian fits of the 1 p.e., 2 p.e. and 3 p.e. peaks.}
  \label{fig:correlated_noise}
\end{figure}

After the IV characterization each test site studied the behaviour of the SiPMs in a completely dark environment at LN2 temperature following the same procedure described in~\cite{andreotti2024cryogenic}. Thanks to the apparatus described in section\,\ref{sec:setup}, the acquired waveforms were used to determine the rate and amplitude of single events and thus DCR. A typical example of the 2D plot of the pulse amplitude as a function of the time delay with respect to the previous pulse, exhibiting all the components of the correlated noise (AP and CT), is shown in figure\,\ref{fig:correlated_noise} on the left side. As clearly visible from the temporal distribution of the events, these SiPMs are affected by the so-called burst effect, in which unexpected trains of consecutive pulses appear at a rate of kHz. This phenomenon is better described in two dedicated publications\,\cite{guarise2021newly,Guarise_2025} and is also present in the HPK lot of DUNE SiPMs\,\cite{andreotti2024cryogenic}. On the right part of figure\,\ref{fig:correlated_noise} the amplitude distribution of the events is shown.

\begin{figure}
\centering
  \includegraphics[width=\textwidth]{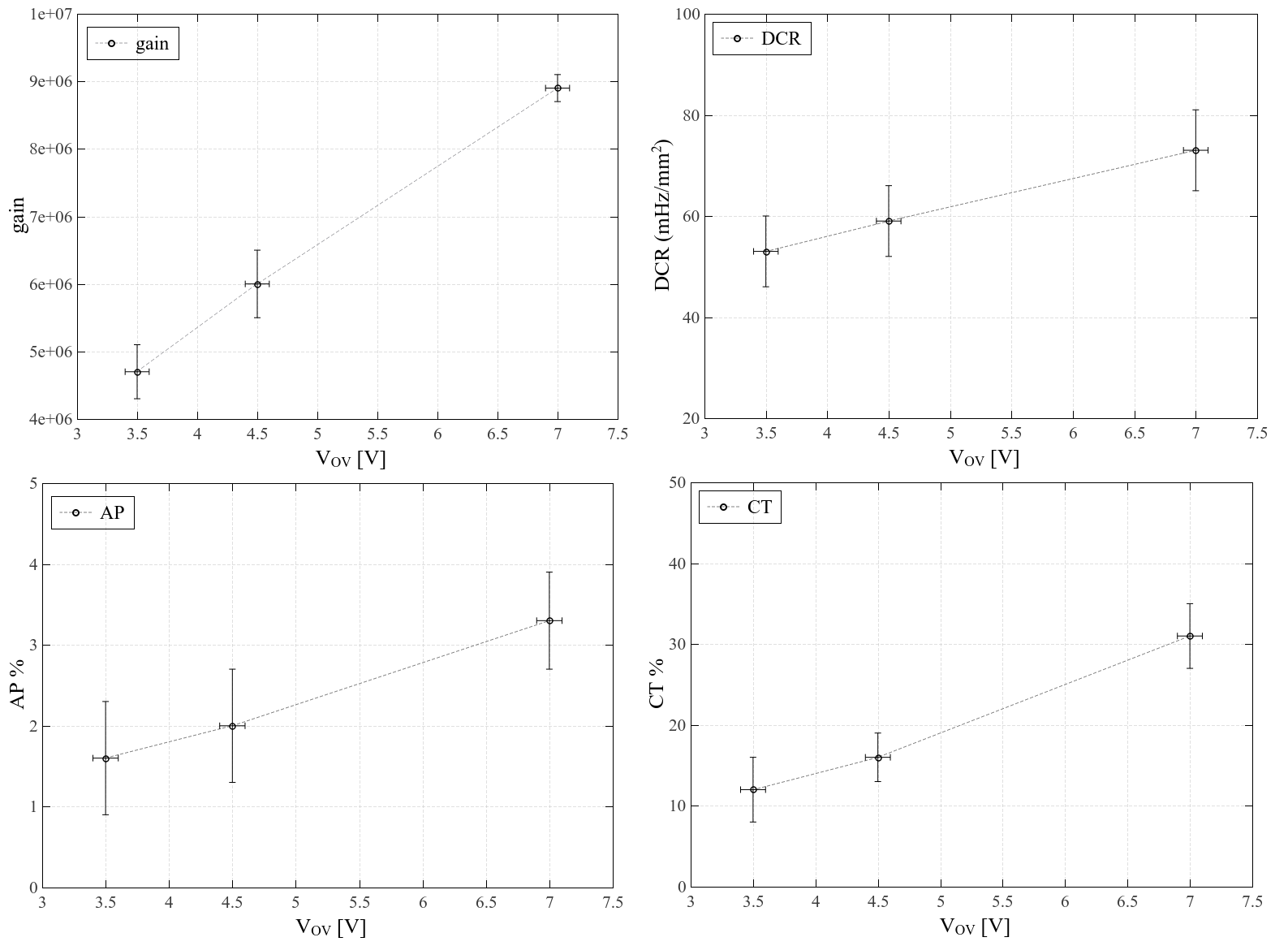}
  \caption{Results of the carachterization of the SiPMs at LN2 temperature. a) Gain as a function of the overvoltage. b) Dark count rate as a function of the overvoltage. c) Afterpulse and d) Crosstalk percentage as a function of the overvoltage.}
  \label{fig:noise_results}
\end{figure}

Figure\,\ref{fig:noise_results} summarizes the main results we obtained for the gain and the correlated noise characterization measured at 3.5\,V, 4.5\,V and 7\,V of overvoltage  that correspond roughly to 40\%, 45\% and 50\% of PDE measured at room temperature as shown in figure\,\ref{fig:FBK_PDE}. We considered the average and the standard deviation of the values obtained independently by the different laboratories. 
The results are in accordance with the values provided by the vendor. The gain spans from $\sim4.5\cdot10^6$ to $\sim9\cdot10^6$ in the range [3.5-7]\,V of overvoltage which is within DUNE requirements.
For the DCR, which is a key parameter in the choice of the photodetectors for the DUNE FDs, we can see that the values are included in the rage [50-80]\,mHz/mm$^2$ for the tested overvoltages. It is worth noting that these values include the burst effect that increase the overall dark counts and is correlated to ionizing radiation that crosses the sensor\,\cite{guarise2021newly}. Since DUNE FDs will be underground, we thus expect a reduction of the burst effect compared to that measured in surface laboratories during the characterization, and consequently a decrease in the global DCR when the SiPMs will be operated in the FDs. However, even with bursts, the measured values for the DCR are within DUNE requirements. The trend of the correlated noise is shown in figures\,\ref{fig:noise_results} c and d for afterpulse and crosstalk effects respectively. All these values are also compliant with the DUNE specifications. The summary of the results is shown in table\,\ref{tab:DCRres}.

\begin{table}[h!]
    \centering
    \begin{tabular}{|c|c|c|c|c|}
    \hline
    \textbf{OV (V)} & \textbf{Gain} &  \textbf{DCR (mHz/mm$^2$)} & \textbf{CT \%} &\textbf{AP \%} \\
    \hline 
    3.5  & (4.7$\pm0.4)\cdot10^6$   &$53\pm7$   &$12\pm4$  &$1.6\pm0.7$\\
    \hline
    4.5& (6.0$\pm0.5)\cdot10^6$     &$59\pm7$   & $16\pm3$ &$2\pm0.7$\\
    \hline
    7 &(8.9$\pm0.2)\cdot10^6$       &$73\pm8$   &$31\pm4$   &$3.3\pm0.6$ \\
    \hline
    \end{tabular}
    \caption{Mean values and standard deviations Gain, DCR and correlated noise obtained from the various laboratories for different overvoltages at LN2 temperature.}
    \label{tab:DCRres}
\end{table}

\section{PDE at cryogenic temperatures}
\label{sec:PDE_meas}

The absolute PDE is a key performance parameter of SiPMs and it is commonly defined as the product of three factors: the SiPM's fill factor, the probability of avalanche triggering, and the silicon's quantum efficiency (QE). The QE measures the likelihood of incident photons generating an electron-hole pair within the sensor’s sensitive volume and depends on photon wavelength. 
At low temperatures the wider energy band gap reduces the quantum efficiency. Furthemore, at approximatively 80-90\,K, carrier freeze-out effects may lead to a decrease in PDE~\cite{Collazuol2011cryogenic}. In Ref. ~\cite{PDE-NUV-Cryo}, a slight decrease in PDE with decreasing temperature was observed for the NUV-HD-Cryo technology.
Therefore, it is essential to directly measure the PDE at the operating temperature (87K in case of liquid argon) for the DUNE customized FBK NUV-HD-Cryo 3T production.\\
\noindent The measurement of the PDE involves characterizing the SiPM response under a calibrated light source. In addition, when determining the PDE, it is crucial to account for the intrinsic SiPMs secondary counts arising from delayed correlated noise (after-pulses and delayed CT).

\subsection{Experimental Setup}
\begin{figure}
\centering
  \includegraphics[width=0.8\textwidth]{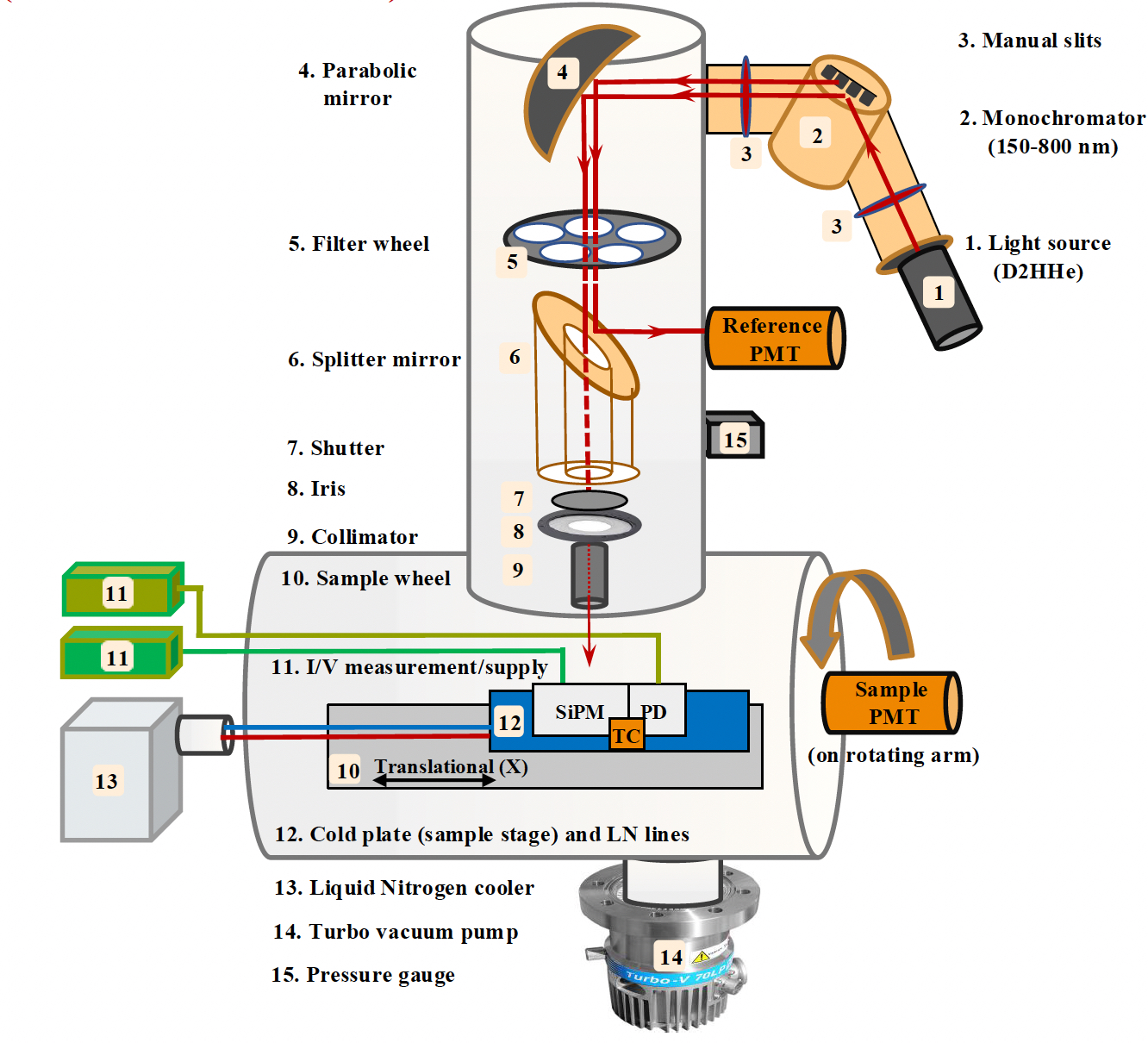}
  \caption{Experimental setup used for PDE measurement of FBK NUV-HD-Cryo 3T sample at 87 K }
  \label{fig:VERA}
\end{figure}

The measurement detailed in this paper were conducted by using the setup called Vacuum Emission Reflectivity Absorbance (VERA) built at TRIUMF in Vancouver to characterize the response of SiPMs at cryogenic temperatures ~\cite{GallinaThesis}.\\
In this setup, the SiPM sample under test was mounted on a cold finger cooled by LN2 and regulated by a control system to maintain the temperature with an accuracy better than 1 K. The setup is consisting of a vacuum chamber coupled to a light source from a Resonance Lyman-Alpha DC Lamp and a VM200 Resonance monochromator. 
In order to measure the absolute incident light flux into the SiPM the light was directed toward a Photo-Diode (AXUV 100G). The AXUV 100G Photo-Diode was previously calibrated, at LAr temperature, against a NIST calibrated Photo-Diode (XUV-100C). The SiPM and the AXUV 100G photodiode are both placed on a movable arm which allows for remote x-y positioning. The incident light selected by monochromator is directed alternatively toward the SiPM under test and the Photo-Diode. In order to avoid geometric factors, the light spot ($1.2$\,mm in diameter) is contained within the area of both sensors.
Throughout the measurement, light stability was monitored by an HPK PMT (R8486) after reflection by a gold coated mirror. The VERA system has already been used to measure the PDE at cryogenic operating temperature in the context of the nEXO (\cite{nEXO1},\cite{nEXO2}) and DarkSide experiments.
A layout of the hardware setup is shown in figure \ref{fig:VERA}.

The SiPM PDE was measured both in photon-counting mode and current mode. In the first case, the SiPM signal was amplified by a dedicated two-stage amplifier and then acquired by a CAEN DT5730B Digitizer Module. The SiPM pulses in this case are individually reconstructed and counted in order to estimate the SiPM rate which will be compared with absolute photon flux measured by Photo-Diode. 
When measured in current mode, a Keysight B2985 low noise picoammeter\footnote{RMS noise with open input of $\sim$140 aA} were used for both the SiPM and the Photo-Diode.  To minimize the noise induced on the Photo-Diode by other sources, a shielded low noise triax cable (Keithley 7078-TRX-1) were used.
A MIDAS-Labview based control system~\cite{MIDAS} served as system interface and provided slow control of the entire hardware system.

\subsection{PDE measurement procedure}
\begin{figure}
\centering
  \includegraphics[width=0.7\textwidth]{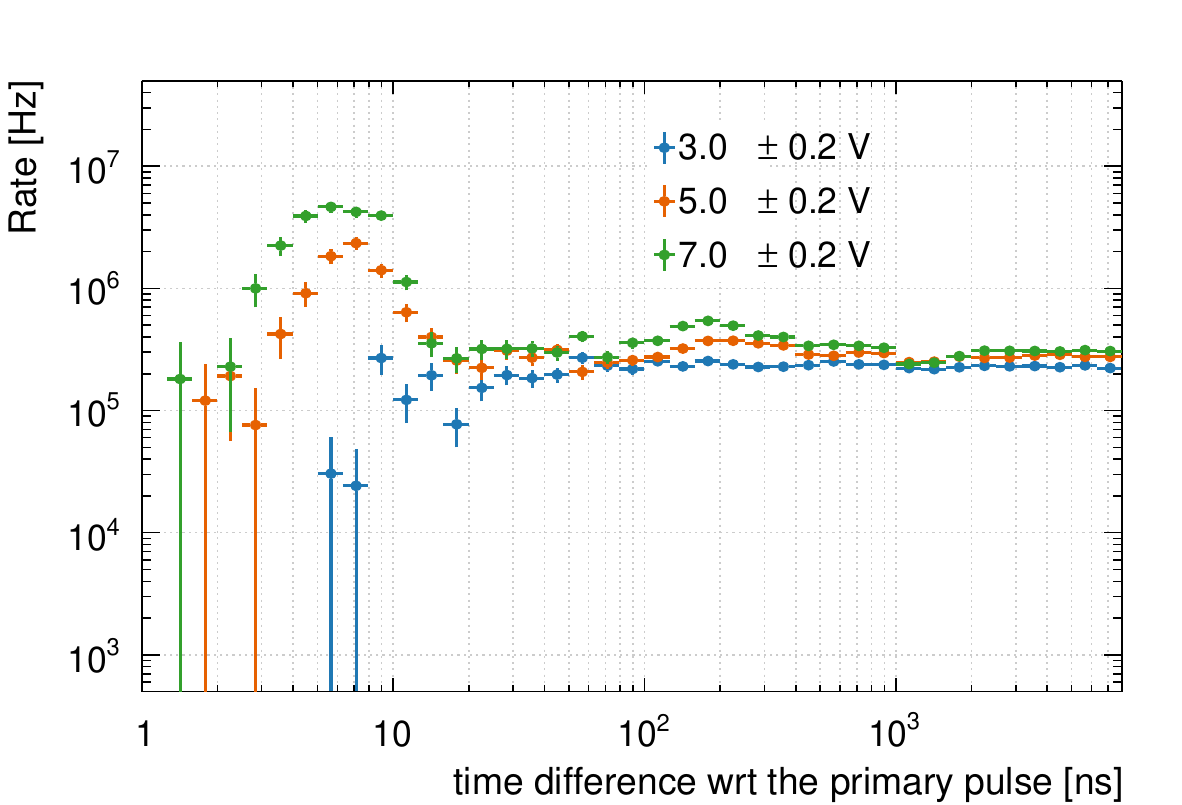}
  \caption{Observed pulse rate R(t) measured at 87K and 420 nm as a function of time differences with respect to the primary pulse for FBK NUV-HD-Cryo 3T sample. The distribution is shown when SiPM is biased at 3, 5, 7 V overvoltages.}
  \label{fig:Rate}
\end{figure}

The first step is to quantify the delayed correlated noise contribution of the SiPM under test as a function of the applied overvoltage. To achieve this, the light wavelength was set to 420 nm and the light level was adjusted to a level suitable to allow simultaneous pulse counting of the individual waveforms and a high enough signal-to-noise ratio of the photodiode to measure the light from the DC lamp. Typically, Photo-Diode currents of the order of $\sim$ 50 fA, were measured in these conditions. Since photons inducing avalanches and dark noise are uncorrelated events, they can be distinguished from correlated delayed avalanches by analyzing the time distribution of all the pulse events relative to the primary pulse, as demonstrated in Ref.~\cite{Butcher} .
The total pulse rate, computed as function of the time difference, $t$, from the primary pulse ($t=0$) is given by:
\begin{equation}\label{eq:Rate}
R(t)=R_{DC}(t)+R_{CDA}(t)+R_{0}(t)
\end{equation}
where $R_{DC}(t)$ is the dark count rate, $R_{CDA}(t)$ is the rate of correlated delayed avalanches per pulse, and $R_{0}(t)$ is the rate of the photon induced avalanche detected by the SiPMs due to light source.
In figure \ref{fig:Rate} the pulse rate R(t) measured at 87 K under 420 nm illumination, is shown as function of the time difference with respect to the primary pulse.  
\noindent As shown in Ref.~\cite{Butcher} the rate of correlated delayed pulses is expected to vanish at sufficiently larger time.  The primary pulse rate $R(t)=R_{DC}(t)+R_{0}(t)$, due to dark count and light source, can then be estimated from the figure \ref{fig:Rate}, performing a weighted mean of the asymptotic rate at long times ($\text{t}>1\times10^3~\text{ns}$).

\noindent Since $R_{DC}$ is independently measured in the absence of light (with the iris diaphragm closed), by using the same method of the rate plot, the true photon-induced rate $R_{0}(t)$ can be measured. The SiPM PDE at 420 $nm$ can be obtained dividing the measured SiPM primary count rate by the the photon flux $\Phi_{0}$ measured by moving the calibrated photodiode under the light beam and it is defined as
\begin{equation}
\label{eq:flux}
\Phi_{0}=\frac{(I-I_{DCR})\lambda}{R(\lambda)hc}
\end{equation}
where $I$ and $I_{DCR}$ are the photodiode currents with and without illumination, respectively, $R$ is the photodiode responsivity at the wavelength $\lambda$, $h$ is Planck’s constant and $c$ is the speed of light. The PDE at 420 nm then is given by

\begin{equation}
\label{eq:pde}
PDE_{420}=\frac{R_{0}(t)}{\Phi_{0}}
\end{equation}

\begin{figure}
\centering
  \includegraphics[width=0.8\textwidth]{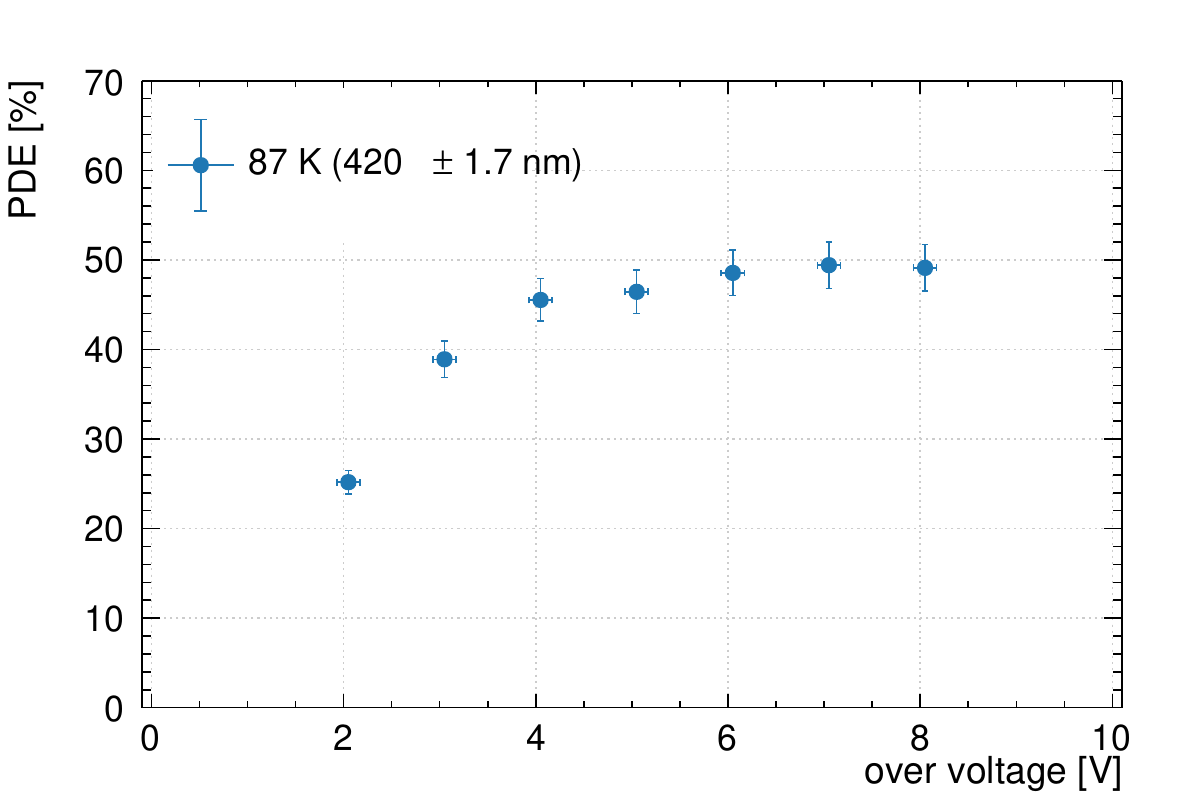}
  \caption{PDE measured with TRIUMF system at 420nm and 87K as a function of the applied overvoltage for the FBK NUV-HD-Cryo 3T sample.}
  \label{fig:PDE420}
\end{figure}

By following this procedure, the quantity $PDE_{420}$ was measured as a function of the applied overvoltage from 2 $V$  up to 8 $V$ (Fig. \ref{fig:PDE420}).

\begin{figure}
\centering
  \includegraphics[width=0.9\textwidth]{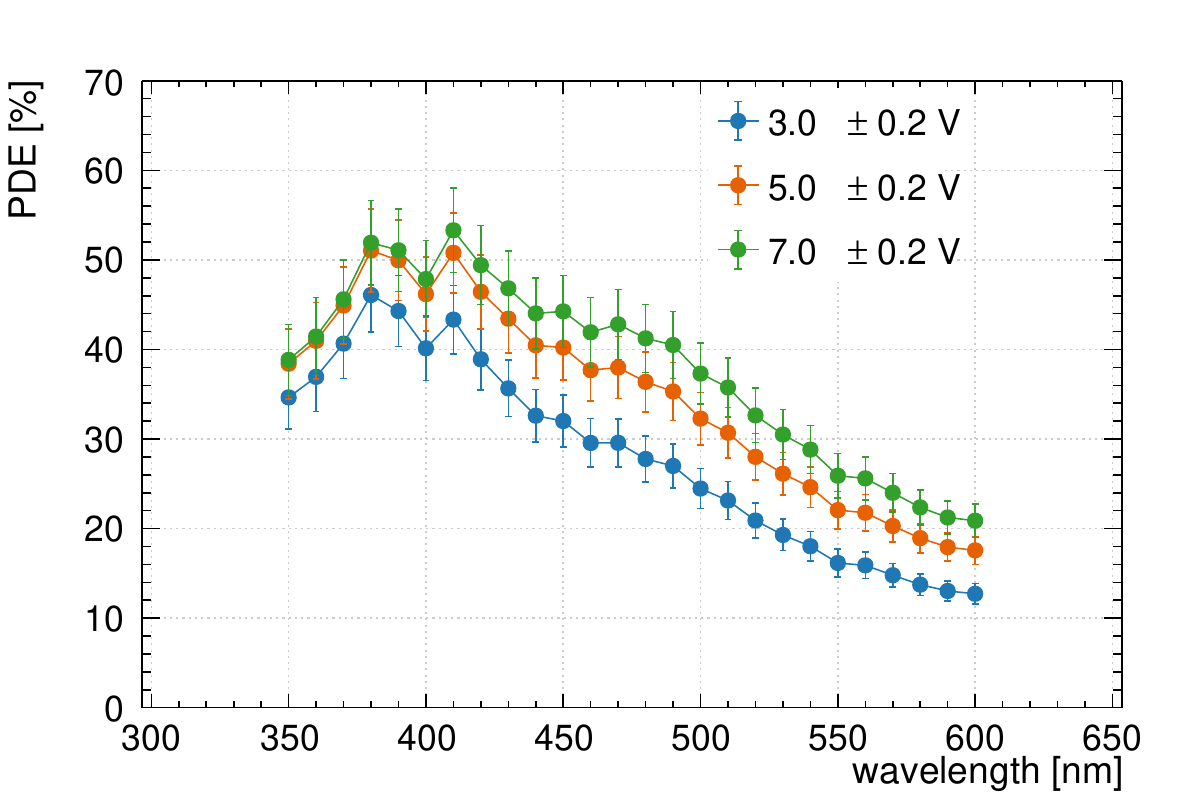}
  \caption{PDE of the FBK NUV-HD-Cryo 3T sample measured with the TRIUMF system at 87 K, as a function of wavelength (350–600 nm) and applied overvoltages of 3, 5, and 7V.}
  \label{fig:PDEScan}
\end{figure}

Measuring the PDE in counting mode is too time-consuming to be performed at every wavelength. The PDE as a function of the wavelength can be determined by measuring the currents of the photodiode and the SiPM scanning wavelengths with the monochromator. Specifically, the SiPM current at a given wavelength $\lambda$ and applied overvoltage $V$ is connected to the PDE through the following formula

\begin{equation}\label{eq:Current}
I_{SIPM}(V,\lambda)-I^{DCR}_{SIPM}(V)=\Phi_{0}(\lambda)\times PDE(V,\lambda) \times f(V)
\end{equation}

\noindent where $I_{SIPM}(V,\lambda)$ and $I^{DCR}_{SIPM}(V)$ are the SiPM current with and without illumination at the wavelength $\lambda$, $\Phi_{0}(\lambda)$ is the photon flux rate measured with photodiode for light illumination at wavelenght $\lambda$, and $f(V)$ is a correction factor that accounts for the SiPM gain and for the correlated avalanche noise, which artifically increase the total SiPM output current. It can be written as
\begin{equation}
\label{eq:fcorr}
f(V) \sim q_e \times (1+ECF) \times G
\end{equation}
where $ECF$ is the extra charge factor and $G$ is the SiPM gain.
It is possible to consider $f(V)$ as a function of the applied bias voltage and to be wavelength-independent because it depends only on the intrinsic characteristics of the SiPM. \\
In this work $f(V)$ is estimated at 87\,K by illuminating SiPMs with continous light source at 420\,nm as follows: 
\begin{equation}
\label{eq:fcorr}
f(V)=\frac{I_{SIPM}(V,420)-I^{DCR}_{SIPM}(V)} {\Phi_{0}(420)\times PDE(V,420)}
\end{equation}
where we take advantage of the fact that the quantity $PDE(V,420)$ was previously measured in counting mode.\\
Once we extracted the wavelength independent factor $f(V)$ for each bias voltage it is possible to measure the PDE for a different wavelength as follows:

\begin{equation}
\label{eq:fcorr}
PDE(V,\lambda)=\frac{I_{SIPM}(V,\lambda)-I^{DCR}_{SIPM}(V)} {\Phi_{0}(\lambda)\times f(V)}
\end{equation}

\noindent where $\Phi_{0}(\lambda)$ is defined in eq.\ref{eq:flux}, $I_{SIPM}(V,\lambda)$ and $I^{DCR}_{SIPM}(V)$ are the SiPM current with and without the illumination at $\lambda$ . Figure \ref{fig:PDEScan} shows the PDE as a function of illumination wavelength, ranging from 350 nm to 600 nm in 10 nm increments, and SIPM overvoltage, varying for 3, 5 and 7 V. The error bars on each point includes statistical and systematic uncertainties. 

\begin{figure}
\centering
  \includegraphics[width=0.9\textwidth]{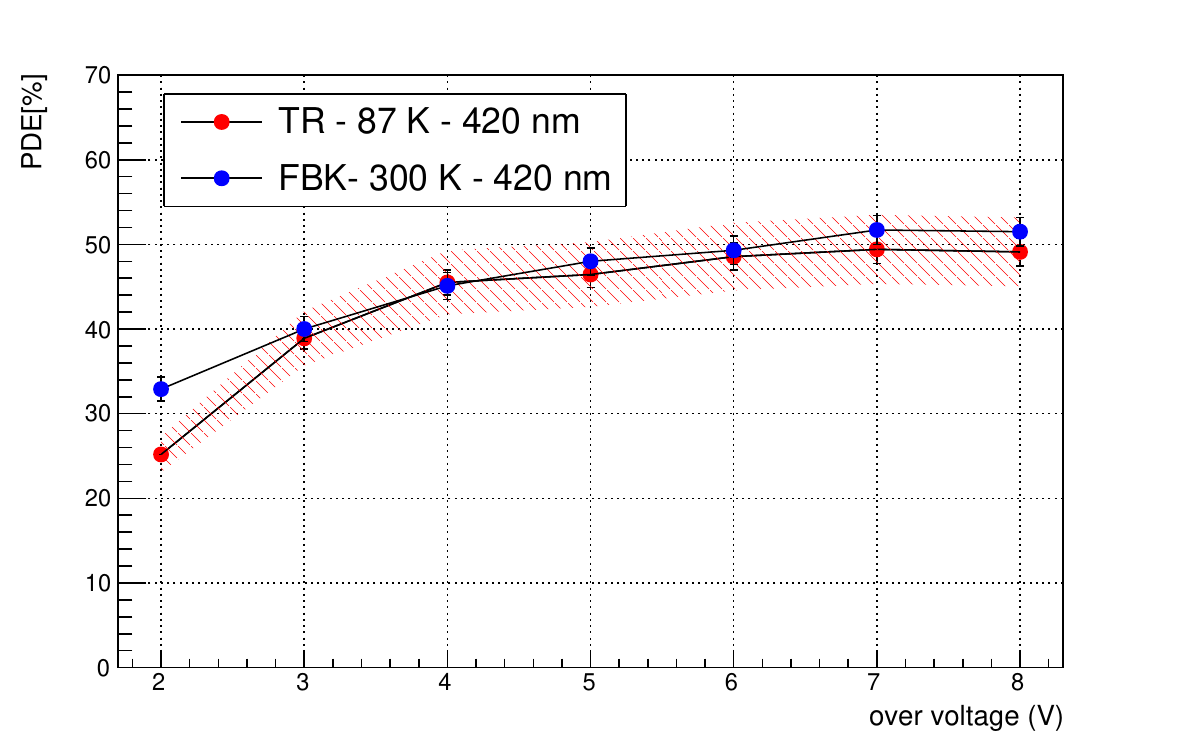}
  \caption{PDE measured at FBK for 420\,nm and 300\,K as a function of the overvoltage (blue dots) and the one measured at TRIUMF for 420\,nm and 87\,K (red dots) for the FBK NUV-HD-Cryo 3T sample. The dashed area represents the region of systematic error in the measurement at 87\,K.}
  \label{fig:PDEComp}
\end{figure}

The PDE measurement were performed at 87\,K temperature. The obtained results are in excellent agreement with the measurements performed at 90\,K, as reported in~\cite{PDE-NUV-Cryo}. We also report in figure \ref{fig:PDEComp} a comparison of the PDE measured at FBK Institute for the same device at room temperature (300\,K) at wavelenght of 420\,nm. 

\section{Conclusions}
\label{sec:conclusion}

In this paper we described the main features and the measurements performed on the NUV-HD-cryo 3T SiPM produced by FBK, to validate this sensor for the DUNE FD-HD PDS. FBK successfully developed this sensor for the DUNE experiment with proprietary technologies aiming a high gain while keeping low the cross-talk probability. The DUNE PDS consortium performed the validation tests of these sensors whose results are shown in this publication. 

The FBK NUV-HD-cryo 3T sensor, together with HPK S13360-9935 were selected for the DUNE FD-HD PDS. Comparing the FBK and the HPK sensors\,\cite{andreotti2024cryogenic}, both provide similar characteristics in terms of gain and correlated noise but the sensor from FBK requires a larger over-voltage for a similar PDE, nevertheless, as the breakdown voltage is lower for the FBK than for the HPK sensors, the operation voltage (for the same PDE) for FBK sensors is lower that the required by the HPK ones.

During the DCR measurements, the appearance of pulse trains (bursts), as also happens with the HPK sensors, increases the DCR value expected at cryogenic temperatures but still meets the DUNE requirements. Those requirements are not very strict because DCR will be dominated by other background sources and achieving the lowest possible single-pe rate is not a critical requirement. Still, a complete understanding of the origin of the DCR burst and the optimal layout of the array field may require further investigation, mainly for experiments with tighter radiogenic constraints.

We also presented a method and the results of the sensor PDE measurement in the wavelengths range from 350\,nm to 600\,nm at liquid argon temperature of 87\,K  in order to validate its performance for the DUNE experiment. Within the measurement errors, no reduction with respect to room temperature PDE was observed.

\acknowledgments

The present research has been supported and partially funded by the Italian Ministero dell’Università e della Ricerca (PRIN 2017KC8WMB and PRIN 20208XN9TZ), by the European Union’s Horizon 2020 Research and Innovation programme under Grant Agreement No 101004761 (AIDAinnova), by the European Union-Next Generation EU by MCIN/AEI/10.13039/501100011033 under Grants no. PID2019-104676GB-C31, No. PID2023-147949NB-C53 \& RYC2022-036471-I and PRE2020–094863 of Spain, by the MEYS of the Czech Republic under grant no. LM2023061, and finally by the University of Ferrara (FIR2023), the BiCoQ Center of the University of Milano Bicocca and  Leonardo Grant for Researchers in Physics 2023 BBVA Foundation. 

We would like to thank the workshop staff from the INFN Ferrara section, namely Alessandro Saputi, Michele Cavallina and Stefano Squerzanti and from the INFN Milano Bicocca section, namely Giancarlo Ceruti, Roberto Mazza, Roberto Gaigher.

\bibliographystyle{JHEP}
\bibliography{biblio.bib} 


\end{document}